\documentclass{article}
\usepackage{graphicx}
\usepackage{psfrag}
\usepackage{flafter}
\def\xslash#1{{\rlap{$#1$}/}}
\begin{document}
\title{Lifetime Difference and Endpoint effect in the Inclusive Bottom Hadron Decays}
\author{Y.B Zuo, Y.A Yan, Y.L Wu and W.Y Wang
\\ \\ Institute of Theoretical Physics, Chinese Academy of
Sciences,\\P.O.Box 2735, Beijing 100080, P.R.China }
\maketitle
\begin{abstract}
The lifetime differences of bottom hadrons are known to be
properly explained within the framework of heavy quark effective
field theory(HQEFT) of QCD via the inverse expansion of the
dressed heavy quark mass. In general, the spectrum around the
endpoint region is not well behaved due to the invalidity of
$1/m_Q$ expansion near the endpoint. The curve fitting method is
adopted to treat the endpoint behavior. It turns out that the
endpoint effects are truly small and the explanation on the
lifetime differences in the HQEFT of QCD is then well justified.
The inclusion of the endpoint effects makes the prediction on the
lifetime differences and the extraction on the CKM matrix element
$|V_{cb}|$ more reliable.
\end{abstract}

\section{Introduction}
The heavy quark effective field theory (HQEFT) of
QCD\cite{ylw1,ylw2} enable us to include in the heavy hadron
decays the nonperturbative binding effects for light degrees of
freedom via introducing the concept of dressed heavy quark. As a
consequence, it has provided a consistent explanation for the
lifetime differences in heavy bottom hadron decays and led to a
reliable extraction on the CKM matrix elements $|V_{cb}|$ and
$|V_{ub}|$ \cite{yww,wy}. Nevertheless the spectrum around the
endpoint region is not well behaved due to the invalidity of
$1/m_Q$ expansion near the endpoint. Especially, an abnormal
behavior that the decay width becomes negative appears in the
endpoint region of the lepton spectrum. References \cite{CWE1,CWE2} give the description of this region using the parton model and its comparison with experiments. In the previous
paper\cite{yww,wy} the endpoint effects on the lifetime
differences are considered to be small. To justify this
consideration, we adopt a curve fitting method to investigate in
detail the endpoint effects. The paper is organized as follows: in
section 2, we present for completeness an explicit evaluation of
the differential decay width. We then compare in section 3 the
lepton spectrum obtained by means of dressed heavy quark expansion
to the next leading order with the one by the free quark model. To
be more clear, we also present the lepton spectrum keeping only
the leading order contribution in the heavy quark expansion of
HQEFT. It is shown that in the endpoint region the contribution
from the next leading order is so large that the expansion doesn't
converge and the expansion method become invalid. To describe the
corresponding physics at endpoint, a curve fitting method is
introduced to treat the endpoint behavior based on the analyticity
and continuity of physics spectrum. Two schemes in the curve
fitting method are considered in order to see how the endpoint
effects influence the observable quantities. Our conclusions are
presented in the last section.

\section{The formulation of lepton spectrum in inclusive semileptonic decays}

Following the evaluation presented in refs.\cite{yww,wy,mw}, the
inclusive decays of heavy hadrons within the framework of
 HQEFT of QCD is described as follows. The semileptonic decays
\begin{eqnarray}
H(P_H=m_Hv) \rightarrow
X_c(P_X)+\ell(p)+\bar{\nu}_{\ell}(p^{\prime})
\end{eqnarray}
can be described by the effective Hamiltonian
\begin{eqnarray}
{\cal H}_{eff}=\frac{4G_f}{\sqrt{2}} V_{cb} \bar{c} \Gamma^{\mu} b \bar{\ell} \Gamma_{\mu} \nu_{\ell}
\end{eqnarray}
To calculate its decay width, we introduce a hadron tensor
\begin{eqnarray}
W^{\mu \nu}=\frac{1}{2 m_{H}} (2 \pi)^3 \sum_{X_c} \delta^4 (P_{H b}-q-P_X) <H|J^{\mu \dagger}|X_c><X_c|J^{\nu}|H>
\end{eqnarray}
where $q=p+p^{\prime}, J^{\mu}=\bar{c} \Gamma^{\mu} b$. Summing
all the contributions with respect to the spin of hadrons in the
initial state and averaging in the final state, according to the
Lorentz invariance, the hadron tensor can be described by five
shape factors:
\begin{eqnarray}
W^{\mu \nu}=-g^{\mu \nu} W_1+v^{\mu} v^{\nu} W_2-i \epsilon^{\mu \nu \alpha \beta} v_{\alpha} q_{\beta} W_3+q^{\mu}
q^{\nu} W_4+(q^{\mu} v^{\nu}+q^{\nu} v^{\mu})W_5
\end{eqnarray}
Then, the differential decay width turns out to be
\begin{eqnarray}
\frac{d \Gamma}{d q^2 d E_{\ell} d E_{\nu}}=\frac{G_{F}^{2} |V_{cb}|^2}{2 \pi^3} \{ [q^2-m_{\ell}^2]W_1+\frac{1}{2}
[4 E_{\ell} E_{\nu}-q^2+m_{\ell}^2] W_2\nonumber \\+[q^2 (E_{\ell}-E_{\nu})-m_{\ell}^2 (E_{\ell}+E_{\nu}) ]W_3
+m_{\ell}^2 \frac{2 E_{\nu}}{\hat{m_b}} W_5 \}
\end{eqnarray}

Furthermore, the shape factors $W_i$ can be evaluated by studying
the following time-ordered product
\begin{eqnarray}
T^{\mu \nu}=- \frac{i}{2 m_H} \int d^4 x e^{- i q \cdot x}<H|{\cal T }\{J^{\mu \dagger}(x) J^{\nu}(0) \}|H>
\nonumber \\ =-g^{\mu \nu} T_1+v^{\mu} v^{\nu} T_2-i \epsilon^{\mu \nu \alpha \beta} v_{\alpha} q_{\beta} T_3+q^
{\mu} q^{\nu} T_4+(v^{\mu} q^{\nu}+v^{\nu} q^{\mu}) T_5
\end{eqnarray}
as the factors $W_i$ are related to $T_i$ via
\begin{eqnarray}
W_i=-\frac{1}{\pi} I_m T_i
\end{eqnarray}
The quark matrix elements in (6) can be represented in the
momentum space as follows
\begin{eqnarray}
\bar{u} \frac{\gamma^{\mu} (m_b {\xslash v}+{\xslash k}-{\xslash q}) \gamma^{\nu}}{(m_b v+k-q)^2-m_{c}^{2}+i \epsilon}
\frac{1-\gamma_5}{2} u
\end{eqnarray}
and the one gluon matrix elements are given by
\begin{eqnarray}
\bar{u} \frac{g}{2} G^{\alpha \beta} \epsilon_{\alpha \beta \kappa \sigma} (m_b v+k-q)^{\kappa} \frac{g^{\mu \sigma}
 \gamma^{\nu}+g^{\nu \sigma} \gamma^{\mu}-g^{\mu \nu} \gamma^{\sigma}+i \epsilon^{\mu \sigma \nu \tau} \gamma_{\tau}}
 {[(m_b v+k-q)^2-m_{c}^{2}+i \epsilon]^2} \frac{1-\gamma_5}{2} u
\end{eqnarray}

From refs. \cite{yww}, we have
\begin{eqnarray}
T_1=\frac{1}{2}(\hat{m}_b-E_{\ell}-E_{\nu})\frac{1}{\Delta}-\frac{1}{2}(3
A+N_b)(\hat{m}_b-E_{\ell}-E_{\nu})\frac{1} {\Delta^2}\nonumber
\\+2 A( \hat{m}_b
-E_{\ell}-E_{\nu})[q^2-(E_{\ell}+E_{\nu})^2]\frac{1}{\Delta^3}
\end{eqnarray}
with $\Delta=(\hat{m}_b v-q)^2-m_c^2+i \epsilon$, $\hat{m}_b
\equiv m_b+v \cdot k$, $A=\frac{\kappa_1}{3}$, $N_b=\frac{d_H
\kappa_2}{3}$. Similarly, $T_2, T_3, T_5$ can all be expanded with
respect to $1/\Delta$.

The lepton spectrum of inclusive semileptonic decay is given
by\cite{yww}
\begin{eqnarray}
\frac{1}{\hat{\Gamma}_0}\frac{d \Gamma}{d y}=-2 \rho^2 (3-\rho)+\frac{4 \rho^3}{(1-y)^3}-\frac{6 \rho^2 (1+\rho)}
{(1-y)^2}+\frac{12 \rho^2}{1-y}+6 (1-\rho) y^2-4 y^3\nonumber \\+\frac{A}{\hat{m}_{b}^2}\{ -6 \rho^3+12 \rho^2
-\frac{24 \rho^3}{(1-y)^5}+\frac{6 \rho^2 (3+5 \rho)}{(1-y)^4}-\frac{12 \rho^2}{(1-y)^3}-\frac{18 \rho^2}{(1-y)^2}
+6 y^2 \}\nonumber \\+\frac{N_b}{\hat{m}_{b}^2}\{ -6 \rho (2+\rho)-\frac{12 \rho^2}{(1-y)^3}+\frac{6 \rho (2+3 \rho)}
{(1-y)^2}-24 \rho y-18 y^2 \}
\end{eqnarray}
with
\[ \hat{\Gamma}_0 \equiv \frac{G_{F}^2 \hat{m}_{b}^5 |V_{cb}|^2}{192
\pi^3}, \qquad y \equiv \frac{2 E_{\ell}}{\hat{m}_b}, \qquad \rho
\equiv \frac{m_c^2}{\hat{m}_b^2}\]

When taking the lepton energy $E_l$ to be in the physics region,
the corresponding region for $y$ is given by
\begin{eqnarray}
0 \leq y \leq
\frac{m_{H_b}}{\hat{m}_b}[1-(\frac{m_{H_c}}{m_{H_b}})^2]
\end{eqnarray}
for light leptons with their masses $m_{\ell}$ being ignored. For
lepton $\tau$, its mass can not be ignored and the corresponding
region is given by
\begin{eqnarray}
2 \sqrt{r_{\tau}} \leq y \leq
\frac{m_{H_b}}
{\hat{m}_b}[1+(\frac{m_{\tau}}{m_{H_b}})^2-(\frac{m_{H_c}}{m_{H_b}})^2]
\end{eqnarray}
with $r_{\tau}=\frac{m_{\tau}^2}{\hat{m}_b^2}$. Here $m_{H_b}$ and
$m_{H_c}$ are the corresponding masses of hadrons containing $b$
quark and $c$ quark respectively.

\section{Endpoint effects and Lifetime difference}

\subsection{Definitions and notations}

The total decay width of bottom hadron $H_b$ is
\begin{eqnarray}
\Gamma_{H_b}^t=\frac{1}{\tau (H_b)} \simeq \Gamma (H_b \rightarrow X_c)+\Gamma (H_b \rightarrow X_{c \bar{c}})+
\sum_{\ell} \Gamma (H_b \rightarrow X_c \ell \bar{\nu})
\end{eqnarray}

To be more precise, the QCD corrections to the decay width should
be considered. The general formulae of the decay width with QCD
corrections can be expressed as follows
\begin{eqnarray}
& & \frac{1}{\hat{\Gamma}_0} \Gamma (H_b \rightarrow X_c \ell
\bar{\nu} )
= \eta_{c \ell} ( \mu ) \Gamma ( \rho )\nonumber \\
& & \frac{1}{\hat{\Gamma}_0} \Gamma (H_b \rightarrow X_c \tau
\bar{\nu} ) =
\eta_{c \tau} ( \mu ) \Gamma ( \rho,\rho_{\tau} )\nonumber \\
& & \frac{1}{\hat{\Gamma}_0} \Gamma (H_b \rightarrow X_c X_{u d} )
= 3 \eta_{c u} ( \mu ) \Gamma ( \rho )
-\frac{1}{\hat{m}_b^2}( c_+^2(\mu) - c_-^2(\mu)) 6 {(1-\rho)}^3 \kappa_2\nonumber \\
& & \frac{1}{\hat{\Gamma}_0} \Gamma (H_b \rightarrow X_c
X_{\bar{c} s} ) = 3 \eta_{c c} ( \mu ) \Gamma ( \rho,\hat{\rho} )
-\frac{1}{\hat{m}_b^2} ( c_+^2(\mu) - c_-^2(\mu) ) I_3 ( \rho ) 6
\kappa_2
\end{eqnarray}
with
\[ \rho_{\tau} \equiv \frac{m_{\tau}^2}{\hat{m}_b^2}, \qquad \hat{\rho}
\equiv \frac{\hat{m}_c^2}{\hat{m}_b^2} \]
 and
\[ c_{\pm}(\mu)={(\frac{\alpha_s(m_W)}{\alpha_s(\mu)})}^{a \pm}, \qquad a_{-}=-2 a_{+}=-\frac{12}{33-2 n_f} \] \\
\[ I_3
(x)=\sqrt{1-4x}(1+\frac{x}{2}+3x^2)-3x(1-2x^2)ln\frac{1+\sqrt{1-4x}}{1-\sqrt{1-4x}}
\]
Using the two loop results obtained in \cite{LS,LL} for the QCD
corrections and taking the input value $m_c/m_b=0.3$, the QCD
correction factors $\eta_{c \ell},\eta_{c \tau},\eta_{c u},\eta_{c
c}$ are given by
\begin{eqnarray}
& & \eta_{c \ell}=1-1.67[\frac{\alpha_s (\mu)}{\pi}]-15.1{[\frac{\alpha_s (\mu)}{\pi}]}^2\nonumber \\
& & \eta_{c \tau}=1-1.39[\frac{\alpha_s (\mu)}{\pi}]-1.58{[\frac{\alpha_s (\mu)}{\pi}]}^2 \beta_0 \\
& & \eta_{cc}=1+2.99[\frac{\alpha_s (\mu)}{\pi}]+ \{ 3.67
\beta_0+3.34 ln (\frac{m_W}{m_b})+4 {[ln (\frac{m_W}
{m_b})]}^2 \}{[\frac{\alpha_s (\mu)}{\pi}]}^2 \nonumber \\
& & \eta_{cu}=1-0.67[\frac{\alpha_s (\mu)}{\pi}]+ \{-1.11
\beta_0+7.17 ln (\frac{m_W}{m_b})+4 {[ln (\frac{m_W} {m_b})]}^2
\}{[\frac{\alpha_s (\mu)}{\pi}]}^2 \nonumber
\end{eqnarray}

 With the above analysis, we are able to calculate the branching ratio of semileptonic decay
\begin{eqnarray}
B_{SL} (H) \equiv B_r (H_b \rightarrow X_c e \bar{\nu})=\frac{\Gamma (H_b \rightarrow X_c e \bar{\nu})}{\Gamma_H^t}
,\nonumber \\
B_{\tau} (H) \equiv B_r (H_b \rightarrow X_c \tau \bar{\nu})=\frac{\Gamma (H_b \rightarrow X_c \tau \bar{\nu})}
{\Gamma_H^t}\nonumber
\end{eqnarray}
the charm counting
\begin{eqnarray}
n_c (H) \equiv 1+\frac{\Gamma (H_b \rightarrow X_{c \bar{c}})}{\Gamma_H^t}-\frac{\Gamma_{nocharm}}{\Gamma_H^t}
\simeq 1+\frac{\Gamma (H_b \rightarrow X_{c \bar{c}})}{\Gamma_H^t}\nonumber
\end{eqnarray}
and the ratio between the $\tau$ and $\beta$ decay :
\begin{eqnarray}
R (H) \equiv \frac{B_{\tau}(H)}{B_{SL}(H)}\nonumber
\end{eqnarray}

 Note that the dressed heavy quark mass $\hat{m}_Q$ is related to the
hadron mass via
\begin{eqnarray}
\hat{m}_Q=m_H\ [1+\frac{\kappa_1-d_H
\kappa_2}{m_H^2}+O(\frac{1}{m_H^3}) ]
\end{eqnarray}
 They differ starting from the order of $1/m_H^2$.

\subsection{The endpoint behavior of the lepton spectrum in heavy quark expansion}

 It is known that the heavy quark expansion becomes unreliable in
 the endpoint region. To see it explicitly, we plot in Fig. 1 the
lepton spectrum of B meson decay via heavy quark expansion in
HQEFT of QCD. For comparison, we also plot the corresponding
spectrum in free quark model. Here we have ignored the lepton
mass. The other relevant parameters are taken to be: $m_c=1.65
Gev,\kappa_1=-0.55{Gev}^2$ and $m_b=4.7$ Gev (for the free quark
model). In general, the numerical calculations involve, besides
the heavy meson and baryon masses, four theoretical parameters.
They are the pole mass $m_c$ of charm quark, the energy scale
$\mu$, and the two hadronic parameters $\kappa_1$ and $\kappa_2$.
Their values are taken to be the same as the ones used
in\cite{yww,wy} in our numerical calculations of this paper.
\begin{eqnarray}
& & m_{B^0}=5.2792 G e v\ ;\ m_{D^+}=1.8693 G e v\nonumber
\\
& & m_{\Lambda_c}=2.2849 G e v \ ;\ m_{\Lambda_b}=5.624 G e v\ ;\ m_{\tau}=1.777 G e v \\
& & 0.3 {Gev}^2 \leq - \kappa_1 \leq 0.7 {Gev}^2 \\
& & \kappa_2 =\frac{1}{8}( m_{B^{\ast 0}}^2-m_{B^0}^2)=0.06 {Gev}^2 \\
& & 1.55 Gev \leq m_c \leq 1.75 Gev
\end{eqnarray}

 From Fig.1, it is easily seen that the decay width in free
quark model (dash-dotted line) is smaller than the one obtained by
using the heavy quark expansion. The reason is simple that in the
former case the nonperturbative binding effects of quarks within
the hadron have been ignored. In the latter case, we have plotted
two curves which correspond to two cases: the leading order
approximation (dashed line) and the next-to-leading order
approximation (solid line). It is seen that when including the
next-to-leading order contributions in the heavy quark expansion,
the decay width becomes negative in the endpoint region of the
lepton spectrum. This is because the lepton energy becomes too
large so that the heavy quark expansion can not be used in this
region. To be more clear, we plot in Fig.2 the corrections of the
next-to-leading order relative to the leading order. On the other
hand, the zero value of the differential decay width is not at the
physics endpoint. This happens also in the free quark model. It
implies that the simple quark-hadron duality is not complete.

 For $\Lambda_b$ baryon decays, the situation is different,
there is no abnormal behavior near the endpoint of the lepton
spectrum. This can be seen explicitly in Fig.3. But it shows that
near the endpoint the next-to-leading order corrections become
larger than the leading order contributions. For the same reason,
the differential decay width does not reached zero
at the physics endpoint.

  It is then clear that we need to smooth the endpoint behavior
and make the spectrum come to zero at the physics endpoint.

\subsection{Curve fitting method and Improvement of endpoint behavior}

   As analyzed above, the heavy quark expansion becomes invalid
in the endpoint region of lepton spectrum as the expansion does
not converge in this region. In the bottom hadron decays, the
abnormal behavior occurs near the endpoint. To be specific, we
define the endpoint region as the region ranges from the point
$y_n$ where the next leading order corrections are one third of
the leading order to the physics endpoint $y_e$. In the endpoint
region, there is no satisfactory approach to treat its behavior in
a reasonable way. In the present paper, we shall adopt a curve
fitting method to treat the endpoint behavior.

To see how the endpoint effects rely on the fitting schemes, we
consider two cases. Firstly, we may consider a simple treatment as
the fitting scheme 1. In this fitting scheme, to avoid the
abnormal behavior in the B meson decays, at the point $y_n$, we
simply replace the next-to-leading spectrum by the leading order
spectrum with multiplying a rescaling factor. Here the rescaling
factor is determined by the continuity condition at the connection
point of the two regions. It is plotted in the figure 4,5. For
$\Lambda_b$ decays, as there is no abnormal behavior, we simply
keep the initial spectrum.

 To ensure all the differential decay widths vanishing at the
physics endpoint, we then introduce a smooth fitting function
\begin{eqnarray}
F_e(y)=\int^{y_e-y}_0 f(x) d x , \qquad
f(x)=e^{a-0.1[\frac{1}{x}+\frac{b}{(1-x)^2}]}
\end{eqnarray}
where $a$, $b$ are two parameters. Choosing appropriate values for them, we
can make the function to be zero at the physics endpoint and to be
equal to unit in the region where the heavy quark expansion is
reasonable so that the lepton spectrums multiplied by these
functions do not change except in the endpoint region.

To be specific, we let the fitting function begin to
slightly reduce at the point where the next leading order corrections are $30\%$ of the leading order. For different decays modes, it needs to
adjust the two parameters to satisfy corresponding conditions. The typical values corresponding to $m_c=1.65Gev,\kappa_1=-0.55{Gev}^2$ are listed explicitly here. \\
For $B^0$ decays:
\begin{eqnarray}
& & m_{\ell}=0: a=25.40; b=185.7 \\
& & m_{\ell}=m_{\tau}: a=34.96; b=270.8 \\
& & m_{\ell}=\hat{m}_c: a=33.56; b=258.2
\end{eqnarray}
For $\Lambda_b$ decays:
\begin{eqnarray}
& & m_{\ell}=0: a=140.70; b=1263 \\
& & m_{\ell}=m_{\tau}: a=299.47; b=2796 \\
& & m_{\ell}=\hat{m}_c: a=416.12; b=3932
\end{eqnarray}
The behavior of the fitting function for the lepton spectrum in
Fig.5 is plotted in Fig.6. Once multiplying the lepton spectrum by
the corresponding fitting function, the behavior of the spectrum
in the endpoint region is well improved and its value at the
physics endpoint is ensured to be zero. This can be seen
explicitly from Fig. 7.

  We now consider another fitting scheme, i.e., scheme 2. Here we adopt the same criterion for $B$ meson and $\Lambda_b$ baryon. Let $y_m$ denote the
point where the spectrum becomes maximum. We divided the spectrum
into three regions $R_m = (0, y_m)$, $R_n = (y_m, y_n)$ and $R_e
=(y_n, y_e)$. We now treat the three regions separately. In the
first region $R_m$ the spectrum is well behaved as the heavy quark
expansion is thought to be reliable.

 In the second region $R_n$, we may introduce a fitting function $F_n(y)$
\begin{eqnarray}
F_n(y)=\alpha + \beta (y-\gamma)^2
\end{eqnarray}
with $ y \geq \gamma$ and $\beta < 0$. The three
parameters $\alpha$, $\beta$ and $\gamma$ are determined by the following conditions:
\begin{eqnarray}
F_n(y_m)=\frac{1}{\hat{\Gamma}_0} \frac{d \Gamma}{d
y}(y_m)|_{A=\frac{\kappa_1}{3} , N_b=\frac{d_H \kappa_2}{3}}
\end{eqnarray}
and
\begin{eqnarray}
F_n(y_n)& & = \frac{\frac{1}{\hat{\Gamma}_0} \frac{d \Gamma}{d
y}(y_n)|_{A=0, N_b=0} }{ 1 - \frac{ \frac{1}{\hat{\Gamma}_0}
\frac{d \Gamma}{d y}(y_n)|_{A=\frac{\kappa_1}{3} , N_b=\frac{d_H
\kappa_2}{3}} - \frac{1}{\hat{\Gamma}_0} \frac{d \Gamma}{d
y}(y_n)|_{A=0, N_b=0} }{ \frac{1}{\hat{\Gamma}_0} \frac{d
\Gamma}{d y}(y_n)|_{A=0, N_b=0} } }\nonumber \\
 & & =\frac{3}{4} \frac{1}{\hat{\Gamma}_0} \frac{d \Gamma}{d
y}(y_n)|_{A=0, N_b=0}
\end{eqnarray}
The typical values for $\alpha$, $\beta$, $\gamma$ correspinding to $m_c=1.65Gev, \kappa_1=-0.55{Gev}^2$ are listed explicitly here.\\
For $B^0$ decays:
\begin{eqnarray}
& & m_{\ell}=0: \alpha=0.9724, \beta=-18.34, \gamma=0.6557\\
& & m_{\ell}=m_{\tau}: \alpha=0.4982, \beta=-20.04, \gamma=0.8367\\
& & m_{\ell}=\hat{m}_c: \alpha=0.5449, \beta=-19.76, \gamma=0.8156
\end{eqnarray}
For $\Lambda_b$ decays:
\begin{eqnarray}
& & m_{\ell}=0: \alpha=1.046, \beta=-20.75, \gamma=0.6770 \\
& & m_{\ell}=m_{\tau}: \alpha=0.6025, \beta=-21.98, \gamma=0.8322 \\
& & m_{\ell}=\hat{m}_c: \alpha=0.4858, \beta=-22.63, \gamma=0.8829
\end{eqnarray}

In the third region $R_e$, namely the endpoint region, we adopt
the leading order spectrum with a rescaling factor to ensure the
continuity at the connection point $y_n$. This is shown in Fig.8.
Then multiplying the smooth fitting function $F_e(y)$ defined in scheme 1. For illustration, we have plotted in the figures
9,10 the spectra of semiletonic decays of B meson and $\Lambda_b$
baryon.

Note that in all the figures we have only illustrated the semileptonic
decays with neglecting the lepton masses. Its generalization to
all the processes is straightforward.

It is seen that in the scheme 2 the spectrum is well improved in
the endpoint region. In addition, the curve fitting method in the scheme 2
adopt the same criterion for the $\Lambda_b$ decays as well, so it is expected to
be more reliable than the scheme 1. We shall present a numerical
calculation in next section to see quantitatively the endpoint
effects.

\subsection{Numerical results and endpoint effects}

Now we calculate the observable quantities using the lepton
spectrum obtained above. First of all we should integrate the
lepton spectrum in the whole physical region to calculate $\Gamma$
for all decay channels. Then the radiative corrections are added
according to (15). In the end, predictions for the observable
quantities are obtained according to their definitions. The
results are shown in Table~\ref{i},~\ref{ii},~\ref{iii},~\ref{iv},~\ref{v},~\ref{vi},~\ref{vii},~\ref{viii}\footnote{In the
present paper, we use $\tau(B^0)=1.540 ps$ and $B_{SL} (B^0)=10.48 \%$ in extracting the CKM
matrix element $|V_{cb}|$. }. The choice of
$\mu$ is subjective and we have adjusted it according to experiments.

In the following, we analyze our results by comparing them with
experiments. The experimental values
 of the observable quanities\cite{yww,DG,CL} are :
\begin{eqnarray}
& & \frac{\tau(\Lambda_b)}{\tau(B^0)}=0.79 \pm 0.05\\
& & B_{SL} (B^0)=10.48 \pm 0.5 \%\\
& & B_{SL} (\Lambda_b)=9.0_{-2.8}^{+3.0} \%\\
& & n_c(B^0)=1.17 \pm 0.04\\
& & B_{\tau}(B^0)=2.6 \pm 0.1 \%\\
& & |V_{cb}|=(41.2 \pm 2.0)\times 10^{-3}
\end{eqnarray}

First, let us see the results in scheme 1 given in
Table~\ref{i},~\ref{ii},~\ref{iii},~\ref{iv}.
Table~\ref{i},~\ref{ii} is corresponding to $\mu = \frac{1}{2}m_b$
and $\mu = m_b$ respectively.\footnote{Here, $m_b=4.4Gev$, which
is different from the value in free quark model.} The lifetime
difference $\frac{\tau(\Lambda_b)}{\tau(B^0)}$ in these two tables
all fall in the range of the experimental uncertainties and are
slightly dependent on parameters $m_c, \kappa_1$ and $\mu$. The
semileptonic branching ratio $B_{SL}(B^0)$ is
sensitive to the choice of $\mu$. Low values of $\mu$ are
corresponding to small values of $B_{SL}(B^0)$. The charm counting
$n_c(B^0)$ is compatible with the experimental data. The numerical
values depend on the pole mass of charm quark and the parameter of
$\kappa_1$. For small pole mass of charm quark, the values of the
charm counting $n_c(B^0)$ become slightly large. In
Table~\ref{iii} we considered the case with fixing the
semileptonic branching ratio to be the experimental central value
$B_{SL}(B_0) = 10.48\%$ by varying the parameter $\mu$. It is seen
that for various values of parameters the energy scale $\mu$ is
found to be $\mu = 2.72 \pm 0.34 Gev$ when the semileptonic
branching ratio $B_{SL}(B^0)$ is fixed to $10.48 \%$.  Taking
$\mu=2.70$ GeV, $m_c=1.65Gev$ and $\kappa_1=-0.55{Gev}^2$ as the
typical values, we present in Table 4 the corresponding
predictions for the various observable quantities. Apparently, the
predictions for the semileptonic braching ratio have been improved
remarkably. In addition, the lifetime difference
$\frac{\tau(\Lambda_b)}{\tau(B^0)}$ is much better than
Table~\ref{i},~\ref{ii}.

The results of scheme 2 are presented in
Table~\ref{v},~\ref{vi},~\ref{vii},~\ref{viii}. Where
Table~\ref{v} and ~\ref{vi} are corresponding to
$\mu=\frac{1}{2}m_b$ and $\mu=m_b$ respectively. The predictions
for $\frac{\tau(\Lambda_b)}{\tau(B^0)}$ are slightly larger than
scheme 1 as a whole. When choosing  $\mu=2.72Gev$, $m_c=1.65Gev$
and $\kappa_1=-0.55{Gev}^2$ as the typical values via
$B_{SL}(B^0)$ being fixed to be $B_{SL}(B^0)= 10.48 \%$, the
corresponding predictions for various observable quantities are
given in Table 8.

Comparing the two schemes, the predictions in the scheme 2 are not
too much different from the scheme 1. It implies that the results
are not sensitive to the fitting scheme. This can also be seen
from Figs.9 and 10, where the two spectra are very similar with
each other. As the fitting method of scheme 2 has adopted the same
criterion for both $B$ and $\Lambda_b$ decays, the results in this
scheme are expected to be more reliable. So we may take the
results in scheme 2 as our theoretical predictions. They are given
by

\begin{eqnarray}
& & \frac{\tau(\Lambda_b)}{\tau(B^0)}=0.80 \pm 0.03\\
& & B_{SL}(B^0)=10.52 \pm 0.45 \%\\
& & B_{SL}(\Lambda_b)=11.23 \pm 0.32 \%\\
& & n_c(B^0)=1.20 \pm 0.02\\
& & B_{\tau}(B^0)=2.6 \pm 0.2 \%\\
& & |V_{cb}|=(40.6 \pm 2.2) \times 10^{-3}
\end{eqnarray}

\section{Conclusion}

In the present paper, we adopt a curve fitting method to improve
the abnormal behavior near the endpoint region and to make a more
reliable calculation for the observable quantities. The resulting
prediction for the lifetime difference is
$\tau(\Lambda_b)/\tau(B^0) = 0.80 \pm 0.03$ which is consistent
with the experimental value $0.79 \pm 0.05$. In is of interest to
note that the endpoint effects to the lifetime difference is small
in the framework of HQEFT of QCD. The predicted value for the
lifetime difference is not sensitive to other theoretical
parameters such as the pole mass of charm quark, the parameter
$\kappa_2$ and the running scale $\mu$. For other predictions, the
large uncertainties mainly arise from the running scale $\mu$. The
resulting central values for charm counting $n_c(B^0)$ seems
slightly larger than the experimental data though it is still
consistent within the errors. The resulting prediction for the CKM
matrix element $|V_{cb}|$ agrees well with the one extracted from
the exclusive decays\cite{wwy}.

In summary, it has been shown that the endpoint behavior may be
smoothed by a curve fitting method. Their contributions are in
general small. The considerations and numerical predictions in the
previous papers\cite{yww,wy} within the framework of HQEFT of QCD
are justified to be reliable. Here the inclusion of the endpoint
effects makes the prediction on the lifetime differences and the
extraction on the CKM matrix element $|V_{cb}|$ more
precise.

 \centerline{\bf Acknowledgement}

 This work was supported in part by the key projects of Chinese
Academy of Sciences, the National Science Foundation of China
(NSFC).


\begin{figure}
\centering
\psfrag{d}[][]{$\frac{1}{\hat{\Gamma}_0} \frac{d
\Gamma}{d y}$} \psfrag{y}[][]{$y$}
\includegraphics{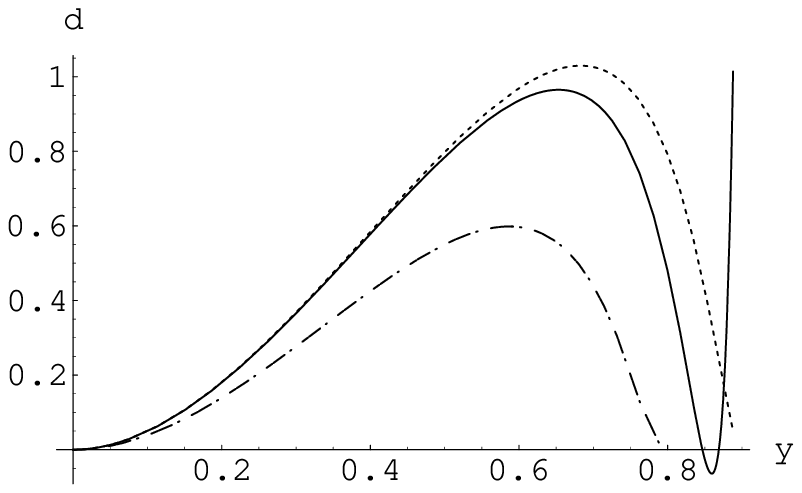}
\caption{Comparison of lepton spectrums in HQEFT with the one in
free quark model. The solid line and the dashed line are
obtained using heavy quark expansion to the next leading order and
to the leading order respectively, whereas the dash dot
line shows the prediction of the free quark model.}
\end{figure}

\begin{figure}
\centering
\psfrag{d}[][]{$\frac{1}{\hat{\Gamma}_0} \frac{d
\Gamma}{d y}$} \psfrag{y}[][]{$y$}
\includegraphics{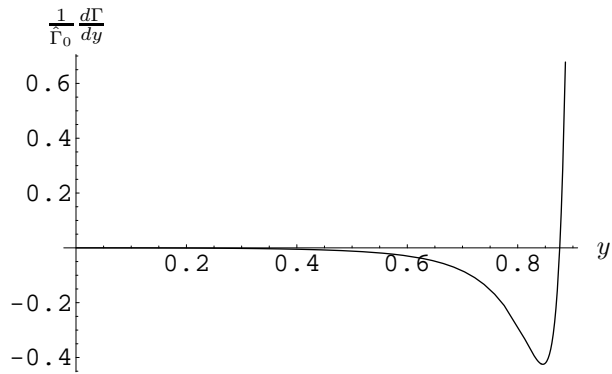}
\caption{The correction of the next leading order to the leading
order of heavy quark expansion in HQEFT.}
\end{figure}

\begin{figure}
\centering
\psfrag{d}[][]{$\frac{1}{\hat{\Gamma}_0} \frac{d \Gamma}{d y}$}
\psfrag{y}[][]{$y$}
\includegraphics{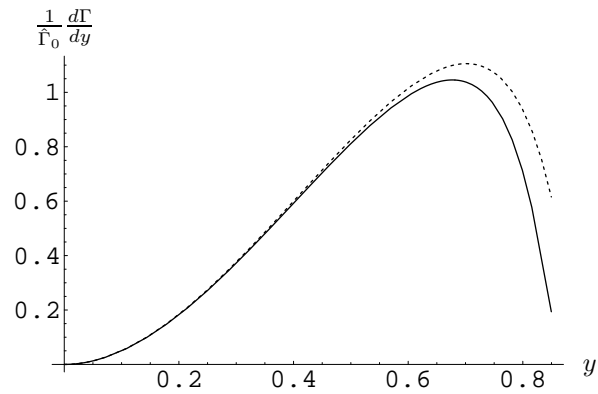}
\caption{The lepton spectrums predicted by heavy quark expansion
in HQEFT corresponding to $\Lambda_b$ decay. The dashed line comes
from the leading order in heavy quark expansion and the solid one
includes the contribution from the next leading order.}
\end{figure}

\begin{figure}
\centering
\psfrag{d}[][]{$\frac{1}{\hat{\Gamma}_0} \frac{d \Gamma}{d y}$}
\psfrag{y}[][]{$y$}
\includegraphics{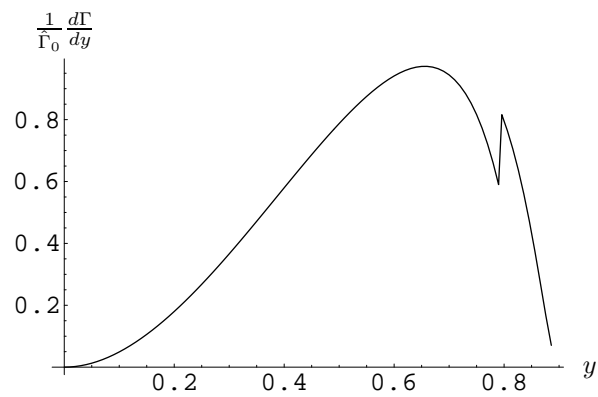}
\caption{The lepton spectrum with the part in endpoint region replaced by the result of leading order.}
\end{figure}

\begin{figure}
\centering
\psfrag{d}[][]{$\frac{1}{\hat{\Gamma}_0} \frac{d \Gamma}{d y}$}
\psfrag{y}[][]{$y$}
\includegraphics{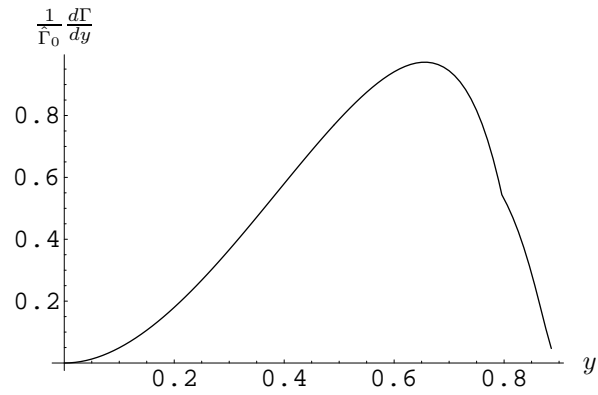}
\caption{The lepton spectrum obtained after multiplying a rescaling factor. }
\end{figure}

\begin{figure}
\centering \psfrag{d}[][]{$F_e ( y )$} \psfrag{y}[][]{$y$}
\includegraphics{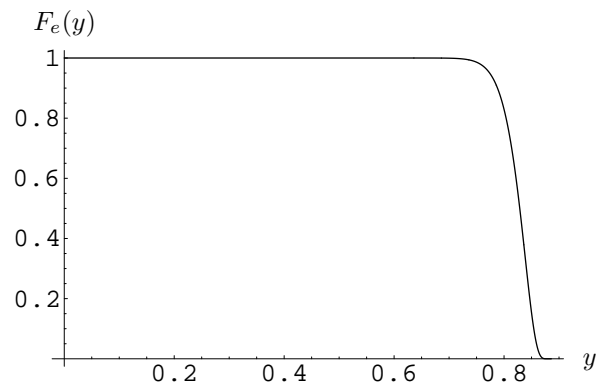}
\caption{The function introduced to make lepton spectrum come to
zero at the end point.}
\end{figure}

\begin{figure}
\centering
\psfrag{d}[][]{$\frac{1}{\hat{\Gamma}_0} \frac{d \Gamma}{d y}$}
\psfrag{y}[][]{$y$}
\includegraphics{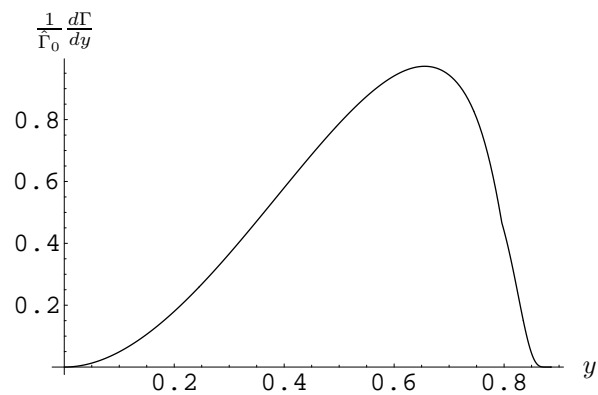}
\caption{The lepton spectrum obtained in scheme 1. }
\end{figure}

\begin{figure}
\centering \psfrag{d}[][]{$\frac{1}{\hat{\Gamma}_0} \frac{d
\Gamma}{d y}$} \psfrag{y}[][]{$y$}
\includegraphics{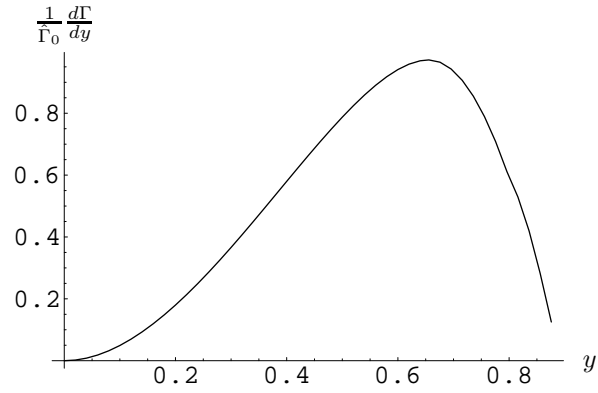}
\caption{The counterpart of Figure 5 in scheme 2.}
\end{figure}

\begin{figure}
\centering \psfrag{d}[][]{$\frac{1}{\hat{\Gamma}_0} \frac{d
\Gamma}{d y}$} \psfrag{y}[][]{$y$}
\includegraphics{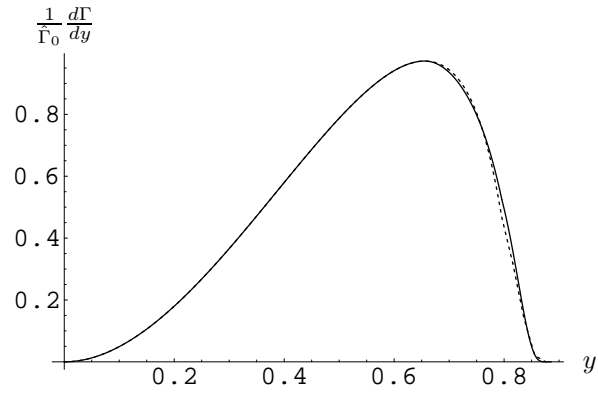}
\caption{The lepton spectrum obtained in scheme 2, where the spectrum in scheme 1 is also given for comparison ( shown in the dashed line ) .}
\end{figure}

\begin{figure}
\centering \psfrag{d}[][]{$\frac{1}{\hat{\Gamma}_0} \frac{d
\Gamma}{d y}$} \psfrag{y}[][]{$y$}
\includegraphics{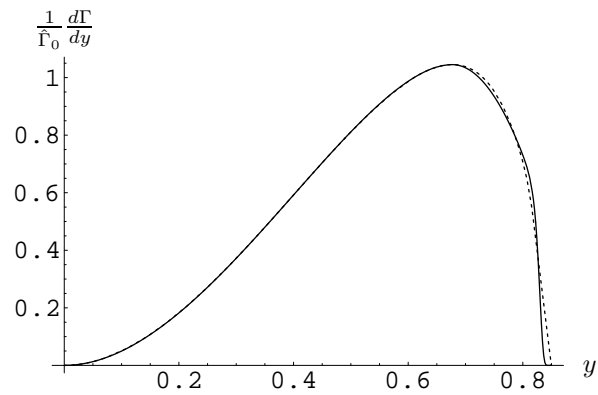}
\caption{The same as Figure 9 but for $\Lambda_b$ decays.}
\end{figure}

\begin{table}
\centering
\tabcolsep0.05in
\begin{tabular}{|c|c|c|c|c|c|c|c|c|c|c|c|c|}
\hline
$m_c (G e v)$ & \multicolumn{3}{|c|}{1.55} & \multicolumn{3}{|c|}{1.65} & \multicolumn{3}{|c|}{1.75} & \multicolumn{3}
{|c|}{1.8}\\
\hline
$\kappa_1 ({G e v}^2)$ & -0.55 & -0.65 & -0.75 & -0.45 & -0.55 & -0.65 & -0.4 & -0.5 & -0.6 & -0.35 & -0.45 & -0.55\\
\hline
$\frac{\tau (\Lambda_b)}{\tau(B^0)}$ & 0.814 & 0.820 & 0.826 & 0.797 & 0.803 & 0.809 & 0.781 & 0.788 & 0.795 & 0.770
 & 0.778 & 0.785\\
\hline
$B_{SL}(B^0)(\%)$ & 9.53 & 9.41 & 9.28 & 9.79 & 9.67 & 9.55 & 10.01 & 9.89 & 9.78 & 10.15 & 10.04 & 9.92\\
\hline
$B_\tau (B^0)$ & 0.026 & 0.025 & 0.024 & 0.026 & 0.025 & 0.023 & 0.025 & 0.024 & 0.023 & 0.025 & 0.023 & 0.022\\
\hline
$R(B^0)$ & 0.27 & 0.27 & 0.26 & 0.27 & 0.26 & 0.24 & 0.25 & 0.24 & 0.24 & 0.25 & 0.23 & 0.22\\
\hline
$n_c(B^0)$ & 1.24 & 1.25 & 1.26 & 1.22 & 1.23 & 1.24 & 1.20 & 1.21 & 1.22 & 1.19 & 1.20 &1.21\\
\hline
$B_{SL}(\Lambda_b)(\%)$ & 10.25 & 10.17 & 10.09 & 10.45 & 10.37 & 10.30 & 10.61 & 10.55 & 10.49 & 10.72 & 10.66 & 10.60\\
\hline
$B_\tau(\Lambda_b)$ & 0.034 & 0.033 & 0.032 & 0.034 & 0.033 & 0.032 & 0.033 & 0.032 & 0.031 & 0.033 & 0.032 & 0.031\\
\hline
$R(\Lambda_b)$ & 0.33 & 0.32 & 0.32 & 0.33 & 0.32 & 0.31 & 0.31 & 0.30 & 0.30 & 0.31 & 0.30 & 0.29\\
\hline
$n_c(\Lambda_b)$ & 1.18 & 1.18 & 1.19 & 1.16 & 1.17 & 1.17 & 1.15 & 1.15 & 1.16 & 1.14 & 1.15 & 1.15\\
\hline
$\frac{B_{SL}(\Lambda_b)}{B_{SL}(B^0)}$ & 1.08 & 1.08 & 1.09 & 1.07 & 1.07 & 1.08 & 1.06 & 1.07 & 1.07 & 1.06 &
1.06 & 1.07\\
\hline
$|V_{cb}|(10^{-2})$ & 3.94 & 4.02 & 4.11 & 4.02 & 4.11 & 4.20 & 4.15 & 4.25 & 4.34 & 4.20 & 4.30 & 4.40\\
\hline
\end{tabular}
\caption{The observable quantities of $B^0, \Lambda_b$ decay in scheme 1, where $\mu=\frac{1}{2} m_b$.} \label{i}
\end{table}

\begin{table}
\centering
\tabcolsep0.05in
\begin{tabular}{|c|c|c|c|c|c|c|c|c|c|c|c|c|}
\hline
$m_c (G e v)$ & \multicolumn{3}{|c|}{1.55} & \multicolumn{3}{|c|}{1.65} & \multicolumn{3}{|c|}{1.75} & \multicolumn{3}
{|c|}{1.8}\\
\hline
$\kappa_1 ({G e v}^2)$ & -0.55 & -0.65 & -0.75 & -0.45 & -0.55 & -0.65 & -0.4 & -0.5 & -0.6 & -0.35 & -0.45 & -0.55\\
\hline
$\frac{\tau(\Lambda_b)}{\tau(B^0)}$ & 0.804 & 0.809 & 0.814 & 0.787 & 0.793 & 0.799 & 0.772 & 0.779 & 0.785 & 0.762
& 0.769 & 0.776\\
\hline
$B_{SL}(B^0)(\%)$ & 11.71 & 11.58 & 11.45 & 12.00 & 11.87 & 11.75 & 12.24 & 12.12 & 12.00 & 12.39 & 12.28 & 12.16\\
\hline
$B_\tau(B^0)$ & 0.031 & 0.030 & 0.029 & 0.031 & 0.030 & 0.028 & 0.030 & 0.029 & 0.027 & 0.030 & 0.028 & 0.027\\
\hline
$R(B^0)$ & 0.26 & 0.26 & 0.25 & 0.26 & 0.25 & 0.24 & 0.25 & 0.24 & 0.23 & 0.24 & 0.23 & 0.22\\
\hline
$n_c(B^0)$ & 1.21 & 1.22 & 1.23 & 1.19 & 1.20 & 1.21 & 1.18 & 1.19 & 1.20 & 1.17 & 1.18 & 1.19\\
\hline
$B_{SL}(\Lambda_b)(\%)$ & 12.46 & 12.38 & 12.30 & 12.68 & 12.60 & 12.53 & 12.85 & 12.80 & 12.73 & 12.97 & 12.91 & 12.85\\
\hline
$B_\tau(\Lambda_b)$ & 0.040 & 0.039 & 0.038 & 0.040 & 0.039 & 0.038 & 0.039 & 0.038 & 0.037 & 0.039 & 0.038 & 0.037\\
\hline
$R(\Lambda_b)$ & 0.32 & 0.32 & 0.31 & 0.32 & 0.31 & 0.30 & 0.30 & 0.30 & 0.29 & 0.30 & 0.29 & 0.29\\
\hline
$n_c(\Lambda_b)$ & 1.15 & 1.16 & 1.17 & 1.14 & 1.15 & 1.15 & 1.13 & 1.13 & 1.14 & 1.12 & 1.13 & 1.13\\
\hline
$\frac{B_{SL}(\Lambda_b)}{B_{SL}(B^0)}$ & 1.06 & 1.07 & 1.07 & 1.06 & 1.06 & 1.07 & 1.05 & 1.06 & 1.06 & 1.05 & 1.05
& 1.06\\
\hline
$|V_{cb}|(10^{-2})$ & 3.70 & 3.78 & 3.86 & 3.78 & 3.86 & 3.94 & 3.90 & 3.99 & 4.08 & 3.95 & 4.04 & 4.13\\
\hline
\end{tabular}
\caption{The observable quantities of $B^0, \Lambda_b$ decay in scheme 1, where $\mu=m_b$.} \label{ii}
\end{table}

\begin{table}
\centering \tabcolsep0.05in
\begin{tabular}{|c|c|c|c|c|c|c|c|c|c|c|c|c|}
\hline $m_c (G e v)$ & \multicolumn{3}{|c|}{1.55} &
\multicolumn{3}{|c|}{1.65} & \multicolumn{3}{|c|}{1.75} &
\multicolumn{3}
{|c|}{1.8}\\
\hline
$\kappa_1 ({G e v}^2)$ & -0.55 & -0.65 & -0.75 & -0.45 & -0.55 & -0.65 & -0.4 & -0.5 & -0.6 & -0.35 & -0.45 & -0.55\\
\hline
$\mu(GeV)$ & 2.83 & 2.94 & 3.06 & 2.61 & 2.70 & 2.80 & 2.46 & 2.54 & 2.62 & 2.38 & 2.44 & 2.52\\
\hline
$\frac{\tau(\Lambda_b)}{\tau(B^0)}$ & 0.810 & 0.815 & 0.820 & 0.794 & 0.800 & 0.805 & 0.779 & 0.786 & 0.792 & 0.769 & 0.776 & 0.783\\
\hline
$B_{\tau}(B^0)$ & 0.028 & 0.027 & 0.026 & 0.027 & 0.026 & 0.025 & 0.026 & 0.025 & 0.024 & 0.025 & 0.024 & 0.023\\
\hline
$R(B^0)$ & 0.27 & 0.26 & 0.25 & 0.26 & 0.25 & 0.24 & 0.25 & 0.24 & 0.23 & 0.24 & 0.23 & 0.22\\
\hline
$n_{c}(B^0)$ & 1.23 & 1.24 & 1.24 & 1.21 & 1.22 & 1.23 & 1.20 & 1.21 & 1.21 & 1.19 & 1.20 & 1.20\\
\hline
$B_{SL}(\Lambda_b)(\%)$ & 11.23 & 11.27 & 11.32 & 11.15 & 11.20 & 11.25 & 11.08 & 11.15 & 11.20 & 11.06 & 11.10 & 11.17\\
\hline
$B_{\tau}(\Lambda_b)$ & 0.037 & 0.036 & 0.035 & 0.036 & 0.035 & 0.034 & 0.034 & 0.034 & 0.033 & 0.034 & 0.033 & 0.032\\
\hline
$R(\Lambda_b)$ & 0.33 & 0.32 & 0.31 & 0.32 & 0.31 & 0.30 & 0.31 & 0.30 & 0.29 & 0.31 & 0.30 & 0.29\\
\hline
$n_c(\Lambda_b)$ & 1.17 & 1.17 & 1.18 & 1.15 & 1.16 & 1.16 & 1.14 & 1.15 & 1.15 & 1.14 & 1.14 & 1.15\\
\hline
$\frac{B_{SL}(\Lambda_b)}{B_{SL}(B^0)}$ & 1.07 & 1.08 & 1.08 & 1.06 & 1.07 & 1.07 & 1.06 & 1.06 & 1.07 & 1.06 & 1.06 & 1.07\\
\hline
$|V_{cb}|(10^{-2})$ & 3.83 & 3.89 & 3.96 & 3.94 & 4.01 & 4.08 & 4.09 & 4.17 & 4.25 & 4.16 & 4.24 & 4.32\\
\hline
\end{tabular}
\caption{The observable quantities of $B^0, \Lambda_b$ decay in scheme 1, where $\mu$ have been chosen to satisfy $B_{SL}(B^0)=10.48 \%$.}
\label{iii}
\end{table}

\begin{table}
\centering
\tabcolsep0.05in
\begin{tabular}{|c|c|c|c|c|c|c|c|c|c|c|c|c|}
\hline
$m_c (G e v)$ & \multicolumn{3}{|c|}{1.55} & \multicolumn{3}{|c|}{1.65} & \multicolumn{3}{|c|}{1.75} & \multicolumn{3}
{|c|}{1.8}\\
\hline
$\kappa_1 ({G e v}^2)$ & -0.55 & -0.65 & -0.75 & -0.45 & -0.55 & -0.65 & -0.4 & -0.5 & -0.6 & -0.35 & -0.45 & -0.55\\
\hline
$\frac{\tau(\Lambda_b)}{\tau(B^0)}$ & 0.811 & 0.816 & 0.822 & 0.793 & 0.800 & 0.806 & 0.778 & 0.785 & 0.792 & 0.767 &
 0.775 & 0.782\\
\hline
$B_{SL}(B^0)(\%)$ & 10.32 & 10.20 & 10.07 & 10.60 & 10.48 & 10.35 & 10.83 & 10.71 & 10.59 & 10.97 & 10.86 & 10.74\\
\hline
$B_{\tau}(B^0)$ & 0.028 & 0.027 & 0.026 & 0.028 & 0.026 & 0.025 & 0.027 & 0.025 & 0.024 & 0.026 & 0.025 & 0.024\\
\hline
$R(B^0)$ & 0.27 & 0.26 & 0.26 & 0.26 & 0.25 & 0.24 & 0.25 & 0.23 & 0.23 & 0.24 & 0.23 & 0.22\\
\hline
$n_c (B^0)$ & 1.23 & 1.24 & 1.25 & 1.21 & 1.22 & 1.23 & 1.19 & 1.20 & 1.21 & 1.18 & 1.19 & 1.20\\
\hline
$B_{SL}(\Lambda_b)(\%)$ & 11.06 & 10.98 & 10.90 & 11.28 & 11.20 & 11.12 & 11.44 & 11.38 & 11.31 & 11.55 & 11.49 & 11.43\\
\hline
$B_\tau(\Lambda_b)$ & 0.036 & 0.035 & 0.034 & 0.036 & 0.035 & 0.034 & 0.035 & 0.034 & 0.033 & 0.035 & 0.034 & 0.033\\
\hline
$R(\Lambda_b)$ & 0.33 & 0.32 & 0.31 & 0.32 & 0.31 & 0.31 & 0.31 & 0.30 & 0.29 & 0.30 & 0.30 & 0.29\\
\hline
$n_c(\Lambda_b)$ & 1.17 & 1.17 & 1.18 & 1.15 & 1.16 & 1.17 & 1.14 & 1.15 & 1.15 & 1.13 & 1.14 & 1.14\\
\hline
$\frac{B_{SL}(\Lambda_b)}{B_{SL}(B^0)}$ & 1.07 & 1.08 & 1.08 & 1.06 & 1.07 & 1.07 & 1.06 & 1.06 & 1.07 & 1.05 & 1.06
& 1.06\\
\hline
$|V_{cb}|(10^{-2})$ & 3.85 & 3.93 & 4.01 & 3.92 & 4.01 & 4.10 & 4.05 & 4.15 & 4.24 & 4.10 & 4.19 & 4.29\\
\hline
\end{tabular}
\caption{The observable quantities of $B^0, \Lambda_b$ decay in scheme 1, where $\mu=2.70Gev$.} \label{iv}
\end{table}

\begin{table}
\centering
\tabcolsep0.05in
\begin{tabular}{|c|c|c|c|c|c|c|c|c|c|c|c|c|}
\hline $m_c (G e v)$ & \multicolumn{3}{|c|}{1.55} &
\multicolumn{3}{|c|}{1.65} & \multicolumn{3}{|c|}{1.75} &
\multicolumn{3}
{|c|}{1.8}\\
\hline
$\kappa_1 ({G e v}^2)$ & -0.55 & -0.65 & -0.75 & -0.45 & -0.55 & -0.65 & -0.4 & -0.5 & -0.6 & -0.35 & -0.45 & -0.55\\
\hline $\frac{\tau(\Lambda_b)}{\tau(B^0)}$ & 0.820 & 0.827 & 0.833
& 0.802 & 0.809 & 0.816 & 0.786 & 0.794 & 0.802 & 0.775 & 0.784 &
0.792\\
\hline $B_{SL}(B^0)(\%)$ & 9.51 & 9.39 & 9.26 & 9.77 & 9.65 & 9.53
& 9.99 & 9.87 & 9.75 & 10.13 & 10.02 & 9.90\\
\hline $B_{\tau}(B^0)$ & 0.026 & 0.025 & 0.024 & 0.026 & 0.025 &
0.024 & 0.025 & 0.024 & 0.023 & 0.025 & 0.024 & 0.023\\
\hline $R(B^0)$ & 0.27 & 0.27 & 0.26 & 0.27 & 0.26 & 0.25 & 0.25 &
0.24 & 0.24 & 0.25 & 0.24 & 0.23\\
\hline $n_c(B^0)$ & 1.24 & 1.25 & 1.26 & 1.22 & 1.23 & 1.24 & 1.20
& 1.21 & 1.22 & 1.19 & 1.20 & 1.21\\
\hline
$B_{SL}(\Lambda_b)(\%)$ & 10.24 & 10.15 & 10.07 & 10.44 & 10.36 & 10.29 & 10.59 & 10.54 & 10.47 & 10.70 & 10.64 & 10.58\\
\hline
$B_\tau(\Lambda_b)$ & 0.034 & 0.033 & 0.032 & 0.034 & 0.033 & 0.032 & 0.033 & 0.032 & 0.031 & 0.033 & 0.032 & 0.031\\
\hline
$R(\Lambda_b)$ & 0.33 & 0.33 & 0.32 & 0.33 & 0.32 & 0.31 & 0.31 & 0.30 & 0.30 & 0.31 & 0.30 & 0.29\\
\hline
$n_c(\Lambda_b)$ & 1.18 & 1.18 & 1.19 & 1.16 & 1.17 & 1.17 & 1.15 & 1.15 & 1.16 & 1.14 & 1.15 & 1.15\\
\hline $\frac{B_{SL}(\Lambda_b)}{B_{SL}(B^0)}$ & 1.08 & 1.08 &
1.09 & 1.07 & 1.07 & 1.08 & 1.06 & 1.07 & 1.07 & 1.06 & 1.06 &
1.07\\
\hline $|V_{cb}|(10^{-2})$ & 3.93 & 4.01 & 4.09 & 4.02 & 4.10 &
4.18 & 4.15 & 4.24 & 4.33 & 4.20 & 4.29 & 4.39\\
\hline
\end{tabular}
\caption{The observable quantities of $B^0, \Lambda_b$ decay in scheme 2, where $\mu = \frac{1}{2} m_b$.} \label{v}
\end{table}

\begin{table}
\centering
\tabcolsep0.05in
\begin{tabular}{|c|c|c|c|c|c|c|c|c|c|c|c|c|}
\hline $m_c (G e v)$ & \multicolumn{3}{|c|}{1.55} &
\multicolumn{3}{|c|}{1.65} & \multicolumn{3}{|c|}{1.75} &
\multicolumn{3}
{|c|}{1.8}\\
\hline
$\kappa_1 ({G e v}^2)$ & -0.55 & -0.65 & -0.75 & -0.45 & -0.55 & -0.65 & -0.4 & -0.5 & -0.6 & -0.35 & -0.45 & -0.55\\
\hline $\frac{\tau(\Lambda_b)}{\tau(B^0)}$ & 0.810 & 0.816 & 0.822
& 0.793 & 0.799 & 0.806 & 0.777 & 0.784 & 0.791 & 0.767 & 0.775 &
0.782\\
\hline $B_{SL}(B^0)(\%)$ & 11.69 & 11.56 & 11.42 & 11.98 & 11.85 &
11.72 & 12.21 & 12.10 & 11.97 & 12.37 & 12.25 & 12.13\\
\hline $B_{\tau}(B^0)$ & 0.032 & 0.030 & 0.029 & 0.031 & 0.030 &
0.029 & 0.030 & 0.029 & 0.027 & 0.030 & 0.028 & 0.027\\
\hline $R(B^0)$ & 0.27 & 0.26 & 0.25 & 0.26 & 0.25 & 0.25 & 0.25 &
0.24 & 0.23 & 0.24 & 0.23 & 0.22\\
\hline $n_c(B^0)$ & 1.21 & 1.22 & 1.23 & 1.19 & 1.20 & 1.21 & 1.18
& 1.19 & 1.20 & 1.17 & 1.18 & 1.19\\
\hline
$B_{SL}(\Lambda_b)(\%)$ & 12.45 & 12.36 & 12.28 & 12.67 & 12.59 & 12.51 & 12.83 & 12.78 & 12.71 & 12.95 & 12.89 & 12.83\\
\hline
$B_\tau(\Lambda_b)$ & 0.041 & 0.040 & 0.038 & 0.040 & 0.039 & 0.038 & 0.039 & 0.038 & 0.037 & 0.039 & 0.038 & 0.037\\
\hline
$R(\Lambda_b)$ & 0.33 & 0.32 & 0.31 & 0.32 & 0.31 & 0.30 & 0.30 & 0.30 & 0.29 & 0.30 & 0.29 & 0.29\\
\hline
$n_c(\Lambda_b)$ & 1.15 & 1.16 & 1.17 & 1.14 & 1.15 & 1.15 & 1.13 & 1.13 & 1.14 & 1.12 & 1.13 & 1.13\\
\hline $\frac{B_{SL}(\Lambda_b)}{B_{SL}(B^0)}$ & 1.07 & 1.07 &
1.08 & 1.06 & 1.06 & 1.07 & 1.05 & 1.06 & 1.06 & 1.05 & 1.05 &
1.06\\
\hline $|V_{cb}|(10^{-2})$ & 3.70 & 3.77 & 3.84 & 3.77 & 3.85 &
3.93 & 3.90 & 3.98 & 4.07 & 3.95 & 4.03 & 4.12\\
\hline
\end{tabular}
\caption{The observable quantities of $B^0, \Lambda_b$ decay in scheme 2, where $\mu = m_b$.} \label{vi}
\end{table}

\begin{table}
\centering \tabcolsep0.05in
\begin{tabular}{|c|c|c|c|c|c|c|c|c|c|c|c|c|}
\hline $m_c (G e v)$ & \multicolumn{3}{|c|}{1.55} &
\multicolumn{3}{|c|}{1.65} & \multicolumn{3}{|c|}{1.75} &
\multicolumn{3}
{|c|}{1.8}\\
\hline
$\kappa_1 ({G e v}^2)$ & -0.55 & -0.65 & -0.75 & -0.45 & -0.55 & -0.65 & -0.4 & -0.5 & -0.6 & -0.35 & -0.45 & -0.55\\
\hline
$\mu(GeV)$ & 2.84 & 2.96 & 3.08 & 2.63 & 2.72 & 2.82 & 2.48 & 2.55 & 2.64 & 2.39 & 2.46 & 2.53\\
\hline
$\frac{\tau(\Lambda_b)}{\tau(B^0)}$ & 0.816 & 0.821 & 0.827 & 0.799 & 0.806 & 0.812 & 0.784 & 0.792 & 0.798 & 0.774 & 0.782 & 0.789\\
\hline
$B_{\tau}(B^0)$ & 0.028 & 0.028 & 0.027 & 0.028 & 0.027 & 0.026 & 0.026 & 0.025 & 0.024 & 0.026 & 0.025 & 0.024\\
\hline
$R(B^0)$ & 0.27 & 0.27 & 0.26 & 0.27 & 0.26 & 0.25 & 0.25 & 0.24 & 0.23 & 0.25 & 0.24 & 0.23\\
\hline
$n_{c}(B^0)$ & 1.23 & 1.24 & 1.25 & 1.21 & 1.22 & 1.23 & 1.20 & 1.21 & 1.22 & 1.19 & 1.20 & 1.21\\
\hline
$B_{SL}(\Lambda_b)(\%)$ & 11.23 & 11.28 & 11.33 & 11.17 & 11.21 & 11.26 & 11.10 & 11.15 & 11.21 & 11.06 & 11.12 & 11.16\\
\hline
$B_{\tau}(\Lambda_b)$ & 0.037 & 0.036 & 0.036 & 0.036 & 0.035 & 0.035 & 0.034 & 0.034 & 0.033 & 0.034 & 0.033 & 0.032\\
\hline
$R(\Lambda_b)$ & 0.33 & 0.32 & 0.32 & 0.32 & 0.31 & 0.31 & 0.31 & 0.30 & 0.29 & 0.31 & 0.30 & 0.29\\
\hline
$n_c(\Lambda_b)$ & 1.17 & 1.17 & 1.18 & 1.15 & 1.16 & 1.17 & 1.15 & 1.15 & 1.15 & 1.14 & 1.14 & 1.15\\
\hline
$\frac{B_{SL}(\Lambda_b)}{B_{SL}(B^0)}$ & 1.07 & 1.08 & 1.08 & 1.07 & 1.07 & 1.07 & 1.06 & 1.06 & 1.07 & 1.06 & 1.06 & 1.06\\
\hline
$|V_{cb}|(10^{-2})$ & 3.82 & 3.88 & 3.94 & 3.93 & 4.00 & 4.07 & 4.09 & 4.16 & 4.24 & 4.15 & 4.23 & 4.31\\
\hline
\end{tabular}
\caption{The observable quantities of $B^0, \Lambda_b$ decay in scheme 2, where $\mu$ have been chosen to satisfy $B_{SL}(B^0)=10.48 \%$.}
\label{vii}
\end{table}

\begin{table}
\centering \tabcolsep0.05in
\begin{tabular}{|c|c|c|c|c|c|c|c|c|c|c|c|c|}
\hline $m_c (G e v)$ & \multicolumn{3}{|c|}{1.55} &
\multicolumn{3}{|c|}{1.65} & \multicolumn{3}{|c|}{1.75} &
\multicolumn{3}
{|c|}{1.8}\\
\hline
$\kappa_1 ({G e v}^2)$ & -0.55 & -0.65 & -0.75 & -0.45 & -0.55 & -0.65 & -0.4 & -0.5 & -0.6 & -0.35 & -0.45 & -0.55\\
\hline $\frac{\tau(\Lambda_b)}{\tau(B^0)}$ & 0.817 & 0.823 & 0.829
& 0.799 & 0.806 & 0.813 & 0.783 & 0.791 & 0.798 & 0.772 & 0.780 &
0.788\\
\hline $B_{SL}(B^0)(\%)$ & 10.33 & 10.20 & 10.07 & 10.60 & 10.48 & 10.36
& 10.83 & 10.71 & 10.59 & 10.98 & 10.86 & 10.74\\
\hline $B_{\tau}(B^0)$ & 0.028 & 0.027 & 0.026 & 0.028 & 0.027 &
0.025 & 0.027 & 0.026 & 0.024 & 0.027 & 0.025 & 0.024\\
\hline $R(B^0)$ & 0.27 & 0.26 & 0.26 & 0.26 & 0.26 & 0.24 & 0.25 &
0.24 & 0.23 & 0.25 & 0.23 & 0.22\\
\hline $n_c(B^0)$ & 1.23 & 1.24 & 1.25 & 1.21 & 1.22 & 1.23 & 1.19
& 1.20 & 1.21 & 1.18 & 1.19 & 1.20\\
\hline $B_{SL}(\Lambda_b)(\%)$ & 11.08 & 10.99 & 10.91 & 11.29 &
11.21 & 11.13 & 11.45 & 11.39 & 11.32 & 11.56 & 11.50 & 11.44\\
\hline $B_{\tau}(\Lambda_b)$ & 0.036 & 0.035 & 0.034 & 0.036 &
0.035 & 0.034 & 0.035 & 0.034 & 0.033 & 0.035 & 0.034 & 0.033\\
\hline $R(\Lambda_b)$ & 0.32 & 0.32 & 0.31 & 0.32 & 0.31 & 0.31 &
0.31 & 0.30 & 0.29 & 0.30 & 0.30 & 0.29\\
\hline $n_c(\Lambda_b)$ & 1.17 & 1.18 & 1.18 & 1.15 & 1.16 & 1.17
& 1.14 & 1.15 & 1.15 & 1.13 & 1.14 & 1.15\\
\hline $\frac{B_{SL}(\Lambda_b)}{B_{SL}(B^0)}$ & 1.07 & 1.08 &
1.08 & 1.07 & 1.07 & 1.07 & 1.06 & 1.06 & 1.07 & 1.05 & 1.06 &
1.07\\
\hline $|V_{cb}|(10^{-2})$ & 3.84 & 3.91 & 3.99 & 3.92 & 4.00 &
4.08 & 4.05 & 4.13 & 4.22 & 4.10 & 4.19 & 4.28\\
\hline
\end{tabular}
\caption{The observable quantities of $B^0, \Lambda_b$ decay in scheme 2, where $\mu = 2.72Gev$.} \label{viii}
\end{table}

\end{document}